\documentclass[graybox]{svmult}
\usepackage{type1cm}        
\usepackage{makeidx}         
\usepackage{graphicx}        
\usepackage[bottom]{footmisc}

\usepackage{newtxtext}       %
\usepackage[varvw]{newtxmath}       

\makeindex             

\begin{document}
	
	\title*{Starving Random Walks}
	\author{ L. R\'egnier\orcidID{0000-0002-3792-1570}, \\ M. Dolgushev \orcidID{0000-0003-3306-9840} and\\ O. B\'enichou \orcidID{0000-0002-6749-2391}}
	\institute{L. R\'egnier \at Sorbonne Universit\'e, Laboratoire de Physique Th\'eorique de la
		Mati\`ere Condens\'ee (LPTMC), 4 Place Jussieu, 75005 Paris, France \email{leo.regnier.pro@outlook.fr}
		\and M. Dolgushev \at Sorbonne Universit\'e, Laboratoire de Physique Th\'eorique de la
		Mati\`ere Condens\'ee (LPTMC), 4 Place Jussieu, 75005 Paris, France \email{maxim.dolgushev@sorbonne-universite.fr}
		\and O. B\'enichou \at Sorbonne Universit\'e, CNRS, Laboratoire de Physique Th\'eorique de la
		Mati\`ere Condens\'ee (LPTMC), 4 Place Jussieu, 75005 Paris, France \email{benichou@lptmc.jussieu.fr}}
	%
	%
	\maketitle
	
	
	\abstract{In this chapter, we review recent results on the starving random walk (RW) problem, a minimal model for resource-limited exploration. Initially, each lattice site contains a single food unit, which is consumed upon visitation by the RW. The RW starves whenever it has not found any food unit within the previous $\mathcal{S}$ steps. \\
	To address this problem, the key observable corresponds to the inter-visit time $\tau_k$ defined as the time elapsed between the finding of the $k^\text{th}$ and the $(k+1)^\text{th}$ food unit. By characterizing the maximum $M_n$ of the inter-visit times $\tau_0,\dots,\tau_{n-1}$, we will see how to obtain the number $N_\mathcal{S}$ of food units collected at starvation, as well as the lifetime $T_\mathcal{S}$ of the starving RW.}
	
\section{Introduction}

When it comes to the exploration of a territory by an animal or biological tracers, a fundamental property in quantifying this exploration is the number $\mathcal{N}(t)$ of sites visited by time $t$ \cite{Alessandro:2021,Gordon:1995}. This quantity has been the focus of many mathematical studies, in which the average, variance, and sometimes the full distribution have been explicitly obtained for a wide range of stochastic processes \cite{Dvoretzky:1951,Jain:1972,Hughes:1995,Gillis:2003,LeGall:1991,Regnier:2022,Vineyard:1963}. These processes include nearest-neighbor jumps on a $d-$dimensional lattice, on fractal media, or RWs with long-range jumps, which model a large variety of biological motions. However, since $\mathcal{N}(t)$ is a cumulative quantity, it does not provide significant information about the past exploration dynamics of the walk: were all the visitations occurring regularly, or were there large time gaps where the RW did not encounter any food units? This differentiation becomes crucial when considering living organisms that rely on finding resources to propel their motion.
	
\begin{figure}[b]
	\centering
	\includegraphics[width=0.7\columnwidth]{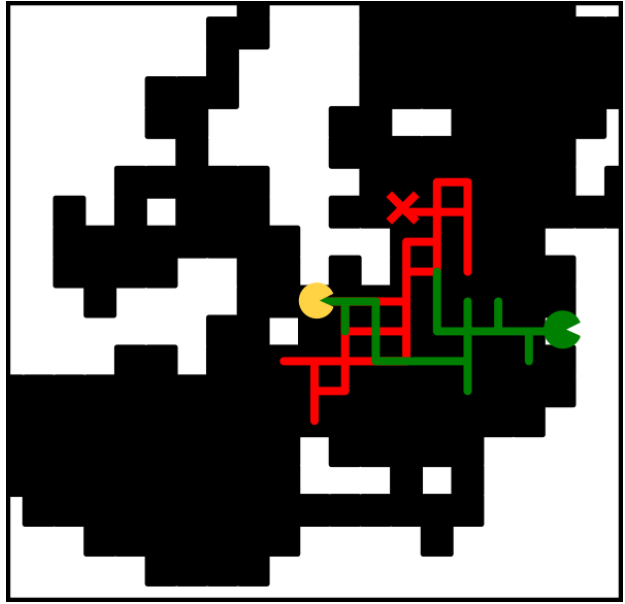}
	\caption{{\bf The starving RW model}. In the lattice, each site initially holds one food unit, which is consumed upon visit by the RW. The RW can make $\mathcal{S}$ steps (here $\mathcal{S}=50$) without encountering food before starving. Two trajectories (green and red) are depicted. A forager (yellow) has consumed $\mathcal{N}(t)=1000$ food units. Areas devoid of food are depicted in black. Along the green trajectory, the forager (green) promptly discovers a new food unit, so that $\tau_{1000}<\mathcal{S}$. Conversely, following the red trajectory, it fails to locate food before succumbing to starvation (depicted by the red cross), $\tau_{1000}\geq \mathcal{S}$.}
	\label{fig:Intro}
\end{figure}

Recently, a new quantity has been introduced to enable this differentiation \cite{Regnier:2022,Benichou:2014,Regnier:2023a,Regnier:2023b,Regnier:2024b}: the time $\tau_k$. This represents the time it takes for a RW to find a site it has never visited before, given that $k$ sites have already been visited. These random variables encapsulate the full dynamics of RW exploration and have been shown to be particularly useful in characterizing the main properties of the starving RW problem \cite{Benichou:2014,Benichou:2016c}, a minimal model to describe depletion-controlled dynamics of a RW \cite{Sowinski:2023,Klages:2023} as encountered in bacteria \cite{Passino:2012}, animals \cite{Orlando:2020}, or robots \cite{Winfield:2009}. 
In the original formulation of the problem, food units are placed on every site of the lattice, and the RW moves through the medium, consuming each food unit upon encounter. The RW starves if $\mathcal{S}$ time steps (the metabolic time) pass without a food encounter (see Fig.~\ref{fig:Intro} for an illustration). This definition establishes a profound connection between $\tau_k$ and starvation: starvation occurs the first time we observe a time between visits to new sites, and thus the finding of a new food unit, larger than the metabolic time. Consequently, understanding the long inter-visit times is fundamental for the search for food, necessitating the study of the extreme-value statistics \cite{Majumdar:2020,Majumdar:2014,Majumdar:2000,ben:2001,kay2023extreme}. We will explore this connection further in the following sections.\\

In this chapter, we examine the visitation properties of general classes of RWs, including anomalous diffusion. We investigate symmetric Markovian RWs that:
\begin{itemize}
	\item Move in a medium of fractal dimension $d_\text{f}$, such that the number of sites within a ball of radius $r$ grows as $r^{d_\text{f}}$.
	\item Have a distance to the origin $r_t$ after $t$ jumps that grows as $t^{1/d_\text{w}}$, where $d_\text{w}$ is the walk dimension \cite{BenAvraham:2000}. More precisely, the distribution of $r_t/t^{1/d_\text{w}}$ is independent of $t$ for large $t$.
\end{itemize}
We note that diffusive RWs in hypercubic lattices of dimension $d$ ($d_\text{w}=2$ and $d_\text{f}=d$), as well as L\'evy flights (jumps whose length distribution $p(\ell)$ is distributed algebraically with exponent $1+\alpha$, $p(\ell)\propto 1/|\ell|^{1+\alpha}$, such that $d_\text{w}=1/\alpha$ for $\alpha<2$) are included in our description.\\

We begin with a brief summary of results on the visitation process of RWs, focusing particularly on the inter-visit time statistics $\tau_k$ of the general RWs under consideration, whose properties depend mainly on the ratio $\mu\equiv d_\text{f}/d_\text{w}$. Then, we delve into the two original cases in which the starving RW problem was examined \cite{Benichou:2014,Benichou:2016c}: the unidimensional and infinite-dimensional cases, which can be viewed as the two limiting scenarios. Finally, we present the general framework for obtaining the properties of both the number $N_\mathcal{S}$ of sites visited at starvation and the lifetime $T_\mathcal{S}$ of the starving RW.
    
\section{Visitation properties}
In this section, we review the results concerning the exploration dynamics of general Markovian processes which will be useful throughout this chapter.
     \subsection{Number of sites visited } \label{sec:Nt}
     In the context of symmetric Markovian RWs characterized by a fractal dimension $d_\text{f}$ and a walk dimension $d_\text{w}$, it was shown in Refs.~\cite{Hughes:1995,Polya:1921,Shepp:1964,Baumler:2023} that there exist two main different classes of exploration: either $\mu\leq 1$, and the RW visits every site infinitely often, or $\mu>1$, and the RW will never visit some sites. The first type of RWs is said to be recurrent, while the second type is said to be transient. Of note, the case $\mu=1$ is differentiated and called marginally recurrent, which we will abbreviate to marginal in the following.\\
     
     This distinction is particularly striking when one considers the average number of sites visited by time $t$ (see Refs.~\cite{Gillis:2003,LeGall:1991,Mariz:2001}),
     \begin{equation} \label{eq:Nt}
     	\left \langle \mathcal{N} (t) \right\rangle \propto \begin{cases}
     		t^\mu & [\mu <1] \\
     		\frac{t}{\ln t} & [\mu=1] \\
     		t & [\mu>1] \; .
     	\end{cases}
     \end{equation}
 
     The difference between the behavior at $\mu<1$ and $\mu>1$ can be understood as follows: while recurrent RWs typically visit the entirety of the ball of volume $r_t^{d_\text{f}}\propto t^\mu$ (the number $t$ of steps they perform is much larger than the number of accessible sites $r_t^{d_\text{f}}=t^{\mu}\ll t$), transient RWs rarely return to previous sites and thus visit a new site at almost every time step. Such strong differences also appear in the exploration dynamics for both types of processes, as we proceed to show in the following section.
     
     \subsection{Inter-visit times statistics} \label{sec:tau_k}
    Below, we summarize the inter-visit time statistics, represented through their probability density $F_k(\tau)\equiv \mathbb{P}(\tau_k=\tau)$, obtained for general Markovian RWs in \cite{Regnier:2023a}.
    \begin{table}[h!]	
	\caption{Summary of the time dependence of $F_k(\tau)$ for the three classes of RWs---recurrent, marginal, and transient. The constants are independent of $k$ and $\tau$. The crossover times $t_k$ and $T_k$ are given up to logarithmic prefactors. }	
	\label{tab:Summary}
	\begin{tabular}{c|cc||ccc}
		\hline\noalign{\smallskip}
	$F_k(\tau)$	& $ t_k $ & $T_k$  & $1 \ll \tau \ll t_k$  & $t_k \ll \tau \ll T_k$  & $T_k \ll \tau $   \\
		\noalign{\smallskip}\svhline\noalign{\smallskip}
		$\mu<1$ [recurrent]& $k^{1/\mu}$ & $k^{1/\mu}$ & &  &    \\
		$\mu=1$ [marginal]& $\sqrt{k}$ & $k^{3/2}$\ & $\tau^{-(1+\mu)}$ & $\exp \Big[ -\text{const}\left(\tau/t_k \right)^{{\mu}/(1+\mu)}\Big]$ &  $\exp  \Big[ -\text{const} \; \tau/k^{1/\mu}\Big]$  \\
		$\mu>1$ [transient] & 1 & $k^{(\mu+1)/\mu}$ & &  &  \\
		\noalign{\smallskip}\hline\noalign{\smallskip}
	\end{tabular}
    \end{table}
    
    The results are as follows:
    \begin{itemize}
    	\item For recurrent RWs, $\mu<1$: there are two distinct time regimes. The first one occurs at times $\tau\ll t_k=k^{1/\mu}$, for which $F_k(\tau)$ decays algebraically with exponent $1+\mu$, independently of $k$. Then, for $\tau \gg t_k$, $F_k(\tau)$  decays exponentially, at a rate of $1/k^{1/\mu}$. Overall, $F_k(\tau)$ admits a scaling form:
    	\begin{equation} \label{eq:scale_rec}
    		F_k(\tau)=\frac{1}{k^{1+\mu}}\psi(\tau/k^{1/\mu})
    	\end{equation}
    	where $\psi$ is a scaling function that depends only on the process.
    	\item For transient RWs, $\mu>1$: there are two distinct time regimes once again, but the early time regime $\tau\ll T_k=k^{1+1/\mu}$ is now characterized by an exponential decay in the variable $\tau^{\mu/(1+\mu)}$ (stretched exponential in $\tau$), and the late time regime is an exponential decay in the variable $\tau/k^{1/\mu}$, as for the recurrent case. There is no algebraic decay.
    	\item For marginal RWs, $\mu=1$: there are now three distinct regimes. Firstly, the distribution $F_k(\tau)$ decays algebraically for $\tau\ll \sqrt{k}$, with an exponent of $2$. This is followed by an exponential decay in the variable $\sqrt{\tau/\sqrt{k}}$ for $\sqrt{k} \ll \tau \ll k^{3/2}$, and finally, an exponential decay for $\tau\gg k^{3/2}$ with a rate of $1/k$. More precisely, in the cross-over regime $\tau \sim \sqrt{k}$, which will be particularly useful in the following, it was shown~\cite{Regnier:2024} that $F_k(\tau)$ admits a scaling form:
    	\begin{equation} \label{eq:scale_marg}
    		F_k(\tau)=\frac{1}{k^{3/2}}\phi(\tau/\sqrt{k})
    	\end{equation}
    	where $\phi$ decays algebraically with an exponent of $3$ at small arguments and exponentially with $\sqrt{\tau/\sqrt{k}}$ at large arguments. With this we get in the cross-over an algebraic decay $\tau^{-3}$ followed by the stretched exponential regime indicated in Table~\ref{tab:Summary}. This rich cross-over behavior comes from the analysis of the exit times from a cascade of the largest visited regions, see Ref.~\cite{Regnier:2024} for details.
    \end{itemize}
    Before applying these fundamental results describing the general $\tau_k$ properties to the starving RW problem, let us start with the two simplest cases were the inter-visit time statistics can be derived exactly.

\section{Low and high dimensional cases} \label{sec:LimitDimStarv}
In this section, we review the two limit cases of unidimensional and infinite dimensional starving RWs, which were solved in \cite{Benichou:2014,Benichou:2016c}.

\subsection{The 1d case}\label{sec:Starv1d}
We start with the case of a symmetric nearest-neighbour RW on a $1d$ line. In this model, the RW jumps with equal probability to one of its two neighbouring lattice sites, on the left or on its right. In this situation, the domain visited by the RW is simple, as it is simply an interval, and the RW starts its new exploration of a new site on the border of the interval (new visits occur only on the left or right border). Thus, the time $\tau_k$ to visit a new site when $k$ sites have already been visited is simply given by the time to exit this interval of length $k$. Consequently, the precise distribution $F_k(\tau)$ of $\tau_k$ is known, and given by the first exit time distribution of the interval starting on one of its border \cite{Redner:2001}:
\begin{equation}
	F_k(\tau)=\frac{2 \pi}{k^2}\sum_{j=0}^\infty (2j+1)\sin\left(\frac{2j+1}{k}\pi\right)\exp\left[ -(2j+1)^2\pi^2\tau/2k^2\right] \; . \label{eq:ExitTime1D}
\end{equation}
One can check that indeed Eq.~\eqref{eq:ExitTime1D} is contained in Table \ref{tab:Summary} for the value $\mu=1/2$, as $d_\text{f}=1$ and $d_\text{w}=2$ (the RW is diffusive). Additionally, we note that the $\tau_k$ are independent, as the distribution of $\tau_k$ only depends on the size $k$ of the visited interval. In the following, we see how the knowledge of the $\tau_k$ provides the distribution of the number $N_\mathcal{S}$ of sites visited at starvation and then of the lifetime $T_\mathcal{S}$.\\

\subsubsection{Number of sites visited at starvation} 
First, we connect the inter-visit time statistics to the starving RW problem using the following remark: to have visited at least $n$ sites before starvation, it means that all the $n$ first inter-visit times $\tau_0$, $\tau_1, \ldots$, $\tau_{n-1}$ are smaller than the metabolic time $\mathcal{S}$. It states that there was no time interval longer than $\mathcal{S}$ between two visits. From this remark, we deduce that:
\begin{equation} \label{eq:NS_tauk_1d}
	\mathbb{P}\left( N_\mathcal{S}\geq n \right)=\mathbb{P}\left(\tau_k < \mathcal{S}, \; k=0, \ldots , n-1 \right) \; .
\end{equation}
From the independence of the $\tau_k$, we set that (in the limit of large $n$ and $\mathcal{S}$):
\begin{equation} \label{eq:Distrib_General}
	\mathbb{P}\left( N_\mathcal{S}\geq n \right)=\prod_{k=0}^{n-1}\left( 1-\sum_{\tau=\mathcal{S}}^\infty F_k(\tau) \right)\sim \exp\left[-\int_{0}^{n}\D k\int_{\mathcal{S}}^\infty\D \tau F_k(\tau) \right] \; .
\end{equation}
Finally, inserting the exact expression of $F_k(\tau)$ given by Eq.~\eqref{eq:ExitTime1D} in Eq.~\eqref{eq:Distrib_General}, we obtain:
\begin{equation} \label{eq:Distrib_NS_1d}
	-\ln \mathbb{P}\left( N_\mathcal{S}\geq n \right)\sim \sum_{j=0}^\infty \int_0^n \frac{4\D k}{k} \E^{-\frac{(2j+1)^2\pi^2\mathcal{S}}{2k^2}} =\sum_{j=0}^\infty \int_1^{\infty} \frac{2\D x}{x} \E^{ -\frac{x(2j+1)^2 \pi^2 \mathcal{S}}{2n^2}}
\end{equation}
where we used that $\sin\left(\frac{2j+1}{k} \pi \right)\sim \frac{2j+1}{k} \pi$ at large $k$. We note that Eq.~\eqref{eq:Distrib_NS_1d} has a scaling form in the variable $\mathcal{S}/n^2$, 
\begin{equation} \label{eq:1D_scale_form}
	\mathbb{P}\left( N_\mathcal{S}\geq n \right)\sim \Xi_{1/2}\left( \frac{\mathcal{S}}{n^2}\right)
\end{equation}
where $\Xi_{1/2}$ is the scaling function, independent of $n$ and $\mathcal{S}$. From Eq.~\eqref{eq:1D_scale_form}, we deduce the scaling with $\mathcal{S}$ of the number of sites visited at starvation:
\begin{eqnarray}
	\left\langle N_\mathcal{S} \right\rangle &\sim& \sum_{n=0}^\infty  \mathbb{P}\left( N_\mathcal{S}\geq n \right) \nonumber \\
	&\sim&  \int_0^\infty \D n \; \Xi_{1/2} \left(\frac{\mathcal{S}}{n^2} \right)=\sqrt{\mathcal{S}}\int_0^\infty \D u \;  \Xi_{1/2} \left(\frac{1}{u^2} \right) \approx 2.9 \sqrt{\mathcal{S}} \; .
\end{eqnarray}

Additionally, one can derive the asymptotic properties of $\Xi_{1/2}(\mathcal{S}/n^2)$ at large and small arguments. For small number of sites visited at starvation 
 $n/\sqrt{\mathcal{S}}\ll 1$, the sum in Eq.~\eqref{eq:Distrib_NS_1d} is dominated by the $j=0$ term, such that:
\begin{equation}
	-\ln \mathbb{P}\left( N_\mathcal{S}\geq n \right) \sim  \int_1^{\infty} \frac{2\D x}{x} \exp\left[ -x\frac{ \pi^2 \mathcal{S}}{2n^2} \right] \sim \frac{4 n^2}{\pi^2 \mathcal{S}}\exp \left[-\frac{\pi^2\mathcal{S}}{2n^2} \right] \; . \label{eq:1dMnShort}
\end{equation}

In the opposite limit of large number of sites visited at starvation $n/\mathcal{S}^2\gg 1$, we use that $F_k(\tau)$ converges to the first passage time distribution $F_\infty(\tau)$ to the site at distance one as the RW does not see the other border of the visited interval. The expression of $F_\infty(\tau)$ is well-known and given by \cite{Redner:2001}:
\begin{equation} \label{eq:Inf_tau_distrib}
	F_\infty(\tau)=\frac{1}{\sqrt{2\pi \tau^3}}\E^{-1/2\tau} \; .
\end{equation}
Once we insert this expression in \eqref{eq:Distrib_General}, we get:
\begin{equation}
	-\ln 	\mathbb{P}\left( N_\mathcal{S}\geq n \right)=\sum_{k=0}^{n-1} \int_{\mathcal{S}}^\infty F_\infty(\tau) \D \tau = n \int_{\mathcal{S}}^\infty \frac{\D \tau }{\sqrt{2\pi \tau^3}}\E^{-1/2\tau} \sim \frac{ n}{\sqrt{8\pi \mathcal{S}}} \; . \label{eq:1dMnLarge}
\end{equation}

\subsubsection{Lifetime} 
We now proceed to determine the lifetime $T_\mathcal{S}$. Initially, we calculate the average lifetime of the RW, which is the sum of inter-visit times before starvation plus the metabolic time, representing the final search for a food unit. This can be expressed as:
\begin{equation}
	\left\langle T_\mathcal{S} \right\rangle =\left\langle \sum_{k=1}^{N_\mathcal{S}} \tau_k \right\rangle +\mathcal{S} =\sum_{n=1}^\infty \left( \sum_{k=1}^n \left\langle \tilde{\tau}_k \right\rangle \right)\mathbb{P}\left(N_\mathcal{S}=n \right)+\mathcal{S} \label{eq:TS_av_formula_1d}
\end{equation}
Here, $\tilde{\tau}_k$ denotes the $k^\text{th}$ inter-visit time conditioned to be less than the metabolic time $\mathcal{S}$. The average of the conditional inter-visit time is given by:
\begin{equation}
	\left\langle \tilde{\tau}_k \right\rangle = \frac{\int_0^{\mathcal{S}} \tau F_k(\tau) \D \tau}{\int_0^{\mathcal{S}} F_k(\tau) \D \tau } \label{eq:Av_cond_intervisit_1D}
\end{equation}

Considering the limit $k\to \infty$ for large visited intervals, $\tau_k$ follows an asymptotic distribution noted $F_{\infty}$, representing the time to exit a semi-infinite interval starting one unit length from the border as given in Eq.~\eqref{eq:Inf_tau_distrib}. Consequently, the limit as $k\to \infty$ of Eq.~\eqref{eq:Av_cond_intervisit_1D} is asymptotically:
\begin{equation}
	\left\langle \tilde{\tau}_\infty \right\rangle = \frac{\int_0^{\mathcal{S}} \tau F_\infty(\tau) \D \tau}{\int_0^{\mathcal{S}} F_\infty(\tau) \D \tau} \sim \sqrt{\frac{2}{\pi}} \sqrt{\mathcal{S}} \; .
\end{equation}
Utilizing Eq.~\eqref{eq:TS_av_formula_1d}, we obtain:
\begin{eqnarray}
	\left\langle T_\mathcal{S} \right\rangle -\mathcal{S} &\sim& \sum_{n=1}^\infty \left( \sum_{j=1}^n \left\langle \tilde{\tau}_\infty \right\rangle \right)\mathbb{P}\left(N_\mathcal{S}=n \right) =\left\langle \tilde{\tau}_\infty \right\rangle\sum_{n=1}^\infty n \mathbb{P}\left(N_\mathcal{S}=n \right) \nonumber \\
	&=&\left\langle \tilde{\tau}_\infty \right\rangle \left\langle N_\mathcal{S} \right \rangle \approx 2.3 \mathcal{S} \; .
\end{eqnarray}

We can demonstrate more generally that the $n^\text{th}$ cumulant of the lifetime $T_\mathcal{S}$ scales as $\mathcal{S}^n$, implying that the random variable $T_\mathcal{S}/\mathcal{S}$ is independent of $\mathcal{S}$ for large $\mathcal{S}$. Denoting by $\kappa_n\left( \ldots\right)$ the $n^\text{th}$ cumulant function ($n>1$), we find that for the $n^\text{th}$ cumulant of the sum of $N_\mathcal{S}$ random variables \cite{Niki:1990}, we have:
\begin{equation}
	\kappa_n\left( T_\mathcal{S} \right)
	\sim \sum_{k_1,\ldots,k_n\in S_n} \frac{n! \kappa_{\sum_{i=1}^n k_i}(N_\mathcal{S})}{k_1!\ldots k_n! (1!)^{k_1}\ldots (n!)^{k_n}}\kappa_1(\tilde{\tau}_\infty)^{k_1}\ldots \kappa_n(\tilde{\tau}_\infty)^{k_n} \label{eq:TS_cum_n_formula_1d}
\end{equation}
where $S_n$ is the set of all vectors $(k_1,\ldots,k_n) \in \mathbb{N}^n$ such that $\sum_{i=1}^n ik_i=n$. Then, from Eq.~\eqref{eq:Inf_tau_distrib}, we find for any $n$,
\begin{equation}
	\left\langle \tilde{\tau}_\infty^n \right\rangle = \frac{\int_0^{\mathcal{S}} \tau^n F_\infty(\tau) \D t}{\int_0^{\mathcal{S}} F_\infty(\tau) \D \tau} \sim \sqrt{\frac{2}{\pi(2n-1)^2}} \mathcal{S}^{n-1/2}
\end{equation}
which implies that $ \kappa_n\left( \tilde{\tau}_\infty \right) \sim \left\langle \tilde{\tau}_\infty^n \right\rangle$. Finally, we obtain that for any vector $(k_1,\ldots,k_n)$ of $S_n$, using that $N_\mathcal{S}$ is scale-invariant from Eq.~\eqref{eq:1D_scale_form} such that, for any $m$, $\kappa_m(N_\mathcal{S})\propto \mathcal{S}^m$,
\begin{eqnarray}
	\kappa_{\sum_{i=1}^n k_i}(N_\mathcal{S})\kappa_1(\tilde{\tau}_\infty)^{k_1}\ldots \kappa_n(\tilde{\tau}_\infty)^{k_n}&&\propto \left( \sqrt{S}\right)^{\sum_{i=1}^n k_i} \mathcal{S}^{k_1(1-1/2)}\ldots \mathcal{S}^{k_n(n-1/2)} \nonumber \\
	&&\propto \mathcal{S}^{\sum_{i=1}^n ik_i}=\mathcal{S}^n
\end{eqnarray}
All the terms in the sum of Eq.~\eqref{eq:TS_cum_n_formula_1d} scale as $\mathcal{S}^n$. This leads to the following scaling with $\mathcal{S}$ for all the cumulants of the lifetime,
\begin{equation}
	\kappa_n\left( T_\mathcal{S} \right) \propto \mathcal{S}^{n} \; .
\end{equation}
In particular, we deduce that the lifetime is scale-invariant, and scales linearly with the metabolic time.
\begin{svgraybox}
	Several important properties stem from the analysis of the $1d$ starving RW:
	\begin{enumerate}
		\item The number of sites visited at starvation grows diffusively with the metabolic time $\mathcal{S}$, $N_\mathcal{S}\propto\sqrt{S}$.
		\item The average lifetime grows linearly with the metabolic time $\mathcal{S}$, $T_\mathcal{S}\propto \mathcal{S}$.  
		\item The distribution of the number of sites visited at starvation and the lifetime are scale-invariant, in the sense that $N_\mathcal{S}/\sqrt{\mathcal{S}}$ and $T_\mathcal{S}/\mathcal{S}$ are non-deterministic and independent of $\mathcal{S}$ for large $\mathcal{S}$. Additionally, the distribution of $N_\mathcal{S}$ decays exponentially at large number of distinct sites visited.
	\end{enumerate}
\end{svgraybox}

These properties are in agreement with what could be expected from the recurrence of the 1D RW (see Sec.~\ref{sec:Nt}): because the RW is recurrent, the starving process stops rapidly, as the RW often comes back to previously visited sites. Besides, we expect the RW not to find a lot of food units before starving. These two observations are in agreement with the linearity of the lifetime with the metabolic time, which is the minimal dependence with $\mathcal{S}$ (as the RW performs at least $\mathcal{S}$ steps before starving). The number of sites visited at starvation corresponds to the domain visited during this short lifetime (see Eq.~\eqref{eq:Nt} for $\mu=1/2$), $N_\mathcal{S} \propto \sqrt{T_\mathcal{S}}\propto \sqrt{\mathcal{S}}$. \\
One might wonder whether these results hold when increasing the dimension. As a first example, the other limit case of infinite dimension was considered in \cite{Benichou:2014,Benichou:2016c}.

\subsection{The infinite dimensional case} \label{sec:StarvInfd}
In this section, we consider the starving RW in a medium of large dimension $d$. A first remark is that in the case of a nearest-neighbor RW in dimension $d$, the constant prefactor in \eqref{eq:Nt} in the transient case can be shown to go to $1$ as the dimension increases (see Refs.~\cite{Watson:1939,Hughes:1995}). This implies that the RW visits a new site at almost every time step, and thus the effect of depletion is small. Consequently, for the RW not visiting a new site, the most likely scenario is to repeat the previous step, returning to its previous position, which occurs with probability $1/(2d)$, a small value. This justifies our assumption that at every time step, the RW visits a new site with probability $\lambda$, independent of time, a value that diminishes in larger dimensions (where $\lambda$ can be interpreted as a generalization of $1/(2d)$ to a RW in a large dimensional space). Additionally,  visitation events are assumed to be independent: they occur rarely, and the visited domain contains almost no loops, as it seldom retraces its past trajectory.

These approximations lead to the effective starving RW model represented in Fig.~\ref{fig:InfDim} (see Ref. \cite{Benichou:2016c}): the effective RW starts at position $\mathcal{S}$, takes one step closer to $0$ with probability $\lambda$, and returns to $\mathcal{S}$ with probability $1-\lambda$ (signifying the finding of a food unit). Similar to the $1d$ case, the derivation begins with determining the number of sites visited at starvation, followed by determining the lifetime.

\begin{figure}[th!]
	\centering
	\includegraphics[width=0.8\columnwidth]{InfiniteDim.pdf}
	\caption{Effective walk representing the starving RW in infinite dimension. The effective RW starts at position $\mathcal{S}$, takes one step closer to $0$ with probability $\lambda$. }
	\label{fig:InfDim}
\end{figure}

\subsubsection{Number of sites visited at starvation}
As a first step, we note that Eq.~\eqref{eq:NS_tauk_1d} holds, as one can reexpress the probability of visiting at least $n$ sites before starving via the statistics of the $\tau_k$, specifically the probability that each of the $\tau_k$ is smaller than $\mathcal{S}$ for $k=0,\ldots, n-1$. The probability that $\tau_k\geq \mathcal{S}$ is exactly given by the probability that the effective RW, represented in Fig.~\ref{fig:InfDim}, arrives at $0$ before returning to $\mathcal{S}$, starting from $\mathcal{S}$ (since the RW just visited a new site). This probability, denoted as $1-P_\mathcal{S}\equiv \mathbb{P}(\tau_k\geq \mathcal{S})$, is independent of $k$ and is given by $1-P_\mathcal{S}=\lambda^\mathcal{S}$, the probability that none of the $\mathcal{S}$ steps lead back to $\mathcal{S}$ (this scaling is in agreement with the $\mu\to \infty$ case of Table \ref{tab:Summary}). Particularly, we note that the $\tau_k$ are again independent, as the state at which any visit starts is the same (the effective RW starts at $\mathcal{S}$).\\

Using Eq.~\eqref{eq:NS_tauk_1d}, we obtain that:
\begin{equation} \label{eq:NS_tau_InfDim}
	\mathbb{P}\left(N_\mathcal{S}\geq n \right)=\prod_{k=0}^{n-1}\mathbb{P}(\tau_k<\mathcal{S})=P_\mathcal{S}^n=(1-\lambda^\mathcal{S})^n \sim \E^{-n\lambda^\mathcal{S}}
\end{equation}
in the limit of large $n$ and small $\lambda$. In particular, we note that the distribution of $N_\mathcal{S}$ is exponential with a rate of $\lambda^\mathcal{S}$, such that:
\begin{equation} \label{eq:NS_av_InfDim}
	\left\langle N_\mathcal{S} \right\rangle \sim 1/\lambda^\mathcal{S} \; .
\end{equation}
Finally, we notice that the distribution of $N_\mathcal{S}/\left\langle N_\mathcal{S} \right\rangle=N_\mathcal{S}\lambda^\mathcal{S}$ is again independent of $\mathcal{S}$, as in the $1d$ case, see Eq.~\eqref{eq:1D_scale_form}.

\subsubsection{Lifetime}
Next, we determine the distribution of the lifetime $T_\mathcal{S}$. To do so, we first compute the probability $F(n,t)$ for the effective RW to arrive at $0$ for the first time at step $t$ starting from position $n$. This distribution obeys a recursive relation, 
\begin{equation} \label{eq:F_n_InfDim}
	F(0,t)=\delta_t , \; F(n,t)=\lambda F(n-1,t-1)+(1-\lambda)F(\mathcal{S},t-1) ,
\end{equation}
where $\delta_t$ is the Kronecker symbol ($1$ if $t=0$, $0$ otherwise). To solve this recursive equation, we Laplace transform Eq.~\eqref{eq:F_n_InfDim}, defined for any $n$ and $s$ by $\hat{F}(n,s)=\mathcal{L} \left( F(n,t) \right)\equiv \sum_{t=0}^\infty F(n,t)\E^{-st}$. This leads to the following equation for $\hat{F}(n,s)$:
\begin{equation}
	\hat{F}(0,s)=1 , \; \E^{-s}\hat{F}(n,s)=\lambda \hat{F}(n-1,s)+(1-\lambda)\hat{F}(\mathcal{S},s)  .
\end{equation}
The solution is given by:
\begin{equation}
	\hat{F}(n,s)=\left( \lambda \E^s \right)^n\left( 1-\frac{\E^s(1-\lambda)}{1-\lambda \E^s} \hat{F}(\mathcal{S},s)\right)+\frac{\E^s(1-\lambda)}{1-\lambda \E^s} \hat{F}(\mathcal{S},s).  \label{eq:F_n_s_lifetime}
\end{equation}
In particular, Eq.~\eqref{eq:F_n_s_lifetime} gives the average lifetime expression, by noting that $\mathbb{P}(T_\mathcal{S}=t)=F(\mathcal{S},t)$:
\begin{equation}
	\left\langle T_\mathcal{S} \right\rangle =\sum_{t=0}^\infty t F(\mathcal{S},t)= -\partial_s \hat{F}(\mathcal{S},s)_{s=0}=\lambda^{-\mathcal{S}}\left( \frac{1-\lambda^\mathcal{S}}{1-\lambda}\right)\sim 1/\lambda^\mathcal{S} \; ,  \label{eq:AvTSInfD}
\end{equation}
which shows that the average lifetime grows exponentially with the metabolic time $\mathcal{S}$. Additionally, we note that in the limit $s/\lambda^\mathcal{S}$ fixed, small $\lambda$ and large metabolic times, 
\begin{equation}
	\mathcal{L} \left( \mathbb{P}(T_\mathcal{S}=t) \right)\sim \frac{\lambda^\mathcal{S}}{s+\lambda^\mathcal{S}}
\end{equation}
so that:
\begin{align}
	\mathbb{P}(T_\mathcal{S}=t) \sim \lambda^\mathcal{S}\E^{-t \lambda^\mathcal{S}}=\frac{1}{\left\langle T_\mathcal{S}\right\rangle}\E^{-t/\left\langle T_\mathcal{S}\right\rangle} \label{eq:InfDimTSdistrib}
\end{align}
which is exponential, and scale-invariant in the sense that $T_\mathcal{S}/\left\langle T_\mathcal{S}\right\rangle$ is a non-deterministic random variable independent of $\mathcal{S}$.

\begin{svgraybox}
Several important properties stem from the analysis of the infinite dimensional starving RW:
\begin{enumerate}
	\item The number of sites visited at starvation grows exponentially with the metabolic time $\mathcal{S}$.
	\item The average lifetime grows exponentially with the metabolic time $\mathcal{S}$.
	\item The distribution of both the number of sites visited at starvation and the lifetime are scale-invariant, in the sense that $N_\mathcal{S} \lambda^\mathcal{S}$ and $T_\mathcal{S} \lambda^\mathcal{S}$ are independent of $\mathcal{S}$ for large $\mathcal{S}$. Additionally, both asymptotic distributions of the rescaled variables decay exponentially at large argument.
\end{enumerate}
\end{svgraybox}
These observations are in agreement with what could be expected for a transient RW: it rarely comes back to previously visited sites, such that the starving process lasts for a long time and finds a lot of sites before starving. This is in agreement with the exponential growth with the metabolic time of both the lifetime and the number of sites visited. \\

Let us now bridge the gap between the one and infinite dimensional cases.

\section{General $d-$dimensional case: the maximum of inter-visit times}
As highlighted in the calculations of the unidimensional and infinite dimensional cases, the derivation of the starving RWs' main observables starts with the determination of the probability that the $n$ first intervisit times are smaller than the metabolic time $\mathcal{S}$, as in Eqs.~\eqref{eq:NS_tauk_1d} and \eqref{eq:NS_tau_InfDim}. The probability of this event can be obtained from the statistics of the maximum $M_n \equiv \underset{k=0,\ldots, n-1}{\max} \tau_k$:
\begin{align} \label{eq:Mn}
	\lbrace \tau_k < \mathcal{S}, \; k=0, \; \ldots , \; n-1 \rbrace=\left\lbrace \max_{k=0,\ldots,n-1}\tau_k < \mathcal{S} \right\rbrace =\lbrace M_n < \mathcal{S} \rbrace \; .
\end{align}

However, the $\tau_k$ are a priori correlated in the general case: indeed, even though the RW is Markovian, values of $\tau_k$ can be representative of different geometries of the visited domain. In the following, we neglect these correlations, because we expect them to be too weak to affect the statistics of the maximum (see \cite{Regnier:2024,Carpentier:2001} for a quantitative argument). This is the reason why we make use of the following approximate expression for the maximum's cumulative distribution function:
\begin{equation} \label{eq:CDF_Mn}
	\mathbb{P}(M_n\leq T)\approx \prod_{k=0}^{n-1}\mathbb{P}(\tau_k\leq T) \sim \exp\left[ -\sum_{k=0}^{n-1}\int_T^\infty F_k(\tau) \D \tau \right] \; . 
\end{equation}

\subsection{Recurrent walks}
For recurrent walks ($\mu<1$), the probability distribution function of $\tau_k$ presents a scaling form (see Eq.~\eqref{eq:scale_rec}), where the scaling function $\psi$ is algebraic at small argument and exponential at large argument (see Table \ref{tab:Summary}). Pluging this information into Eq.~\eqref{eq:CDF_Mn} implies that the cumulative distribution function of $M_n$ also has a scaling form:
\begin{eqnarray}
	- \ln \mathbb{P}(M_n\leq T)&\approx&  \sum_{k=0}^{n-1} \int_T^\infty k^{-1-1/\mu}\psi(\tau/k^{1/\mu}) \D \tau \nonumber \\
	&\sim& \int_{0}^{n/T^\mu} \frac{\D  v}{v}\int_{v^{-1/\mu}}^\infty \psi(u) \D u  =- \ln \Xi_\mu (T/n^{1/\mu})  \; .
	\label{eq:RecScale}
\end{eqnarray}

Let us now consider the limit behaviours of $\Xi_\mu(m)$ when $m \equiv T/n^{1/\mu} \ll 1$ and $ m \gg 1$.\\

First, in the limit $m \gg 1$, using Table \ref{tab:Summary},
\begin{equation}
	-\ln \Xi_\mu (m)\propto \int_0^{1/m^{\mu}} \frac{\D x'}{x'} \E^{-A/x'^{1/\mu}} \propto \int_{m}^\infty \frac{\D x'}{x'} \E^{-A x'} \propto \E^{-Am}/m \; . \label{eq:RecMnLarge}
\end{equation}

Then, in the limit $m \ll 1$
\begin{equation}
	-\ln \Xi_\mu (m) \propto  \text{const.}+ \int_1^{1/m^{\mu}} \frac{\D x'}{x'} x' \propto  m^{-\mu}. \label{eq:RecMnShort}
\end{equation}
We will see the implications of these scalings on the starving RW problem in Sec.~\ref{sec:Starving}. 

\subsection{Marginal walks}
Based on the functional form \eqref{eq:scale_marg} for $F_k(\tau)$, we expect $M_n/\sqrt{n}$ to converge to a non-trivial distribution at large $n$. Indeed, for $m>0$ fixed and $n\to \infty$, we have that:
\begin{eqnarray}
	&&-\ln \mathbb{P}(M_n \leq m \sqrt{n}) \approx  \int_0^n {\rm d } k  \int_{m\sqrt{n}}^\infty \D \tau F_k(\tau) \nonumber  \\
	 &&=\int_0^{m^{2/3}n^{1/3}} {\rm d } k  \int_{m\sqrt{n}}^\infty \D \tau \exp\left[- \text{const. } \tau /k \right] + \int_{m^{2/3}n^{1/3}}^n \frac{ {\rm d } k}{k} \int_{m\sqrt{n}}^\infty \frac{\D \tau}{\sqrt{k}}\phi\left(\frac{\tau}{\sqrt{k}}\right) \nonumber \\
	&&\sim \int_0^{1/m^2} \frac{\D u}{u} \int_{1/\sqrt{u}}^\infty {\rm d } t \phi\left( t \right) =-\ln \Xi_1(m) \; . 
\end{eqnarray}
From this, it is apparent that $M_n/\sqrt{n}$ has an asymptotic limit distribution, denoted by $\Xi_1$. The properties of $\Xi_1(m)$ at small and large $m$ are directly derived from the properties of $\phi$ introduced in Eq.~\eqref{eq:scale_marg} and following similar steps as Eqs.~\eqref{eq:RecMnLarge} and \eqref{eq:RecMnShort}:
\begin{equation} \label{eq:MargDistMn}
	-\ln \Xi_1(m)\propto \begin{cases}
		m^{-2} &\text{ for } m\ll 1 \\
		\E^{-A\sqrt{m}} &\text{ for } m\gg1
	\end{cases}
\end{equation}
where $A$ is a process dependent constant and all algebaric prefactors have been neglected in the $m\gg 1$ regime. 

\subsection{Transient walks}
For transient walks, $\mu>1$, according to Table \ref{tab:Summary}, the decrease of the distribution of $\tau_k$ is first exponential in $\tau^{\mu/(1+\mu)}$ (stretched exponential in $\tau$ of exponent $\delta=\mu/(1+\mu)$), a regime which lasts up to time $T_k=k^{1+1/\mu}$, and next exponential in $\tau/k^{1/\mu}$. On the other hand, the maximum $M_n$ of $n$ random variables with such stretched exponential tail (but no exponential tail) has an average which grows as $\left\langle M_n \right\rangle \sim a \left(\ln n\right)^{1/\delta}=a\left(\ln n \right)^{1+1/\mu}$ ($a$ a model-dependent constant) and a standard deviation which grows as $\sqrt{\text{Var}\left( M_n \right)}=\sqrt{\left\langle M_n^2 \right\rangle-\left\langle M_n \right\rangle^2} \propto \left(\ln n\right)^{1/\delta-1}=\left(\ln n \right)^{1/\mu}$, as shown in Ref.~\cite{Majumdar:2020}.  In particular, this means that the maximum is peaked around its mean (standard deviation much smaller than its mean) which is itself small compared to the time $n^{1+1/\mu}$ (typical value of $T_k$ for $k$ between $0$ and $n-1$) at which the stretched exponential regime would stop. We conclude that the asymptotic distribution of the maximum $M_n$ of the $\tau_k$ is indeed not influenced by the exponential decay at large times. Thus, the limit law (centered and normalized) of $M_n$ corresponds to the celebrated Gumbel distribution \cite{Gumbel:1958} of zero mean and unit variance of cumulative distribution function we note $\Xi_\infty$. For any value of $\mu$, the cumulative distribution function of $M_n$ centered and normalized is given by:
\begin{equation}
	-\ln \Xi_\mu(m)=-\ln \Xi_\infty(m)= \E^{-m\E^{-\gamma_E}}
\end{equation} 
where $\gamma_E$ is the Euler constant and the rescaled random variable $m$ is defined in Eq.~\eqref{eq:def_m}.

\begin{important}{Important}
	By considering the rescaled random variable ($a$ a model-dependent constant)
	\begin{equation}\label{eq:def_m}
		m\equiv \begin{cases}
			M_n/n^{1/\mu} & \; [\mu<1] \\
			M_n/\sqrt{n} & \; [\mu=1] \\
			\frac{M_n-a\left(\ln n\right)^{1/\mu+1}}{\left( \ln n\right)^{1/\mu}} &\; [\mu>1] 
		\end{cases}
	\end{equation}
    we obtain its asymptotic cumulative distribution $\Xi_\mu$, whose behavior at large and small arguments is summarized in Table~\ref{tab:Mn}.
    \begin{table}[h!]
    	\centering	
    	\caption{Asymptotic properties of the cumulative distribution $\Xi_\mu$ of the properly rescaled maximum for different types of processes. All constants have been put to one for simplicity.}	
    	\label{tab:Mn}
    	\begin{tabular}{ccc}
    		\hline\noalign{\smallskip}
    		$-\ln \Xi_\mu(m)$ & $m\ll 1$ & $m\gg1$   \\
    		\noalign{\smallskip}\svhline\noalign{\smallskip}
    		$\mu<1$ [recurrent]& $m^{-\mu}$ &  $\E^{-m}/m$     \\
    		$\mu=1$ [marginal]& $m^{-2}$ &  $\E^{-\sqrt{m}}$   \\
    		$\mu>1$ [transient] & $\E^{-m}$ & $\E^{-m}$  \\
    		\noalign{\smallskip}\hline\noalign{\smallskip}
    	\end{tabular}
    \end{table}
\end{important}

\section{Starving Random Walks}\label{sec:Starving}

We now show that the knowledge of the cumulative distribution function of $M_n$ allows one to determine the statistics of the number of sites visited at starvation and the lifetime of a starving RW.

\subsection{Number of sites visited at starvation}
\begin{figure}[th!]
	\centering
	\includegraphics[width=\columnwidth]{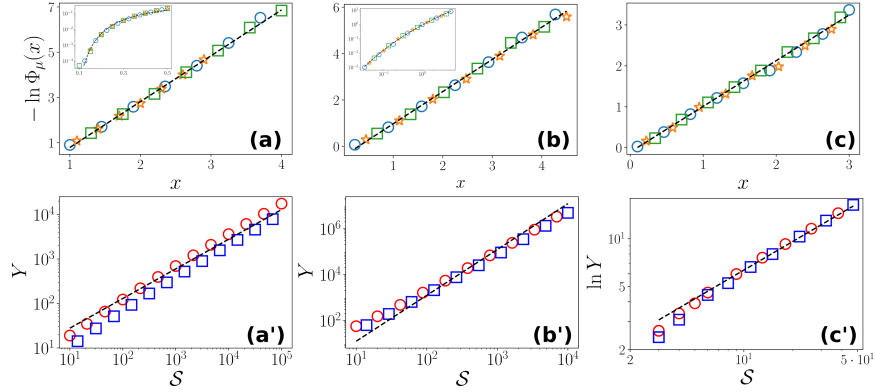}
	\caption{{\bf Number of sites visited at starvation $N_\mathcal{S}$.} {\bf (a)-(c)} Distributions of the rescaled variable $x\equiv N_\mathcal{S}/\langle N_\mathcal{S}\rangle$ (insets illustrate the small $x-$values) and {\bf (a$^\prime$)-(c$^\prime$)} the corresponding averages (red circles) and standard deviations (blue squares) of $N_\mathcal{S}$. The black dashed lines correspond to the best fit of the theory (see Eqs.~\eqref{eq:NSRec}, \eqref{eq:NSMarg}, \eqref{eq:NSTrans}, and \eqref{eq:momNS}). Different universality classes are represented by {\bf (a)} $1d$ L\'evy flights of parameter $\alpha=1.5$ (recurrent), $\mathcal{S} = 14667$, $31622$ and $68129$;  {\bf (b)} nearest neighbour $2d$ RWs, $\mu=1$ (marginal), $\mathcal{S}= 2335$, $4832$ and $10^4$; {\bf (c)} nearest neighbour $3d$ RWs, $\mu=3/2$ (transient), $\mathcal{S}=12$, $17$ and $22$.  Increasing values of $\mathcal{S}$ are represented by blue circles, orange stars, and green squares.}
	\label{fig:DistribN}
\end{figure}

We start from the remark that having $M_n$ smaller than the metabolic time $\mathcal{S}$ is the same as having visited at least $n$ sites before starvation, thus
\begin{equation}
	\mathbb{P}\left(N_\mathcal{S}\geq n \right) =\mathbb{P}\left(M_n<\mathcal{S} \right) \; .
\end{equation}
This means that one can deduce the distribution of $N_\mathcal{S}$ directly from that of $M_n$.
\begin{itemize}
	\item For recurrent RWs: The distribution of $N_\mathcal{S}$ is scale-invariant (so that $N_\mathcal{S}/\mathcal{S}^\mu$ is independent of $\mathcal{S}$) and its asymptotics is given by:
	\begin{equation}
		-\ln\mathbb{P}(N_\mathcal{S}/\mathcal{S}^\mu \geq x)=-\ln \Xi_\mu (x^{-1/\mu}) \propto 
		\begin{cases}
			\E^{-A/x^{1/\mu}}x^{1/\mu} &\mbox{ for } x \ll 1 \\
			x &\mbox{ for } x\gg 1.
		\end{cases}
		\label{eq:NSRec}
	\end{equation}
	Of note, this result is consistent with what we found in Sec.~\ref{sec:Starv1d}, Eqs.~\eqref{eq:Distrib_NS_1d}, \eqref{eq:1dMnLarge} corresponding to $\mu=1/2$.
	\item For marginal RWs: As $\mathbb{P}(N_{\mathcal{S}=\sqrt{n}x}  \geq n)=\mathbb{P}(M_n \leq \sqrt{n}x) \to \Xi_1(x) $, with $\Xi_1(x)$  of Eq.~\eqref{eq:MargDistMn}, we get that ${\mathbb{P}(N_\mathcal{S}/\mathcal{S}^2 \geq x) \to \Xi_1(1/\sqrt{x})}$. In particular,  the distribution has a single-parameter scaling at large starvation index $\mathcal{S}$ and number of visited sites at starvation $N_\mathcal{S}$, and its average and standard deviation scale both as $\mathcal{S}^2$. Based on Eq.~\eqref{eq:MargDistMn}, up to logarithmic corrections,
	\begin{equation}
		\label{eq:NSMarg}
		-\ln \mathbb{P}(N_\mathcal{S}/\mathcal{S}^2 \geq x) \propto
		\begin{cases}
			\E^{-A/x^{1/4}} &\mbox{ for } x\ll 1 \\
			x &\mbox{ for } x\gg 1.
		\end{cases}
	\end{equation}
	\item For transient RWs, considering $\left\langle M_n\right\rangle \sim a (\ln n)^{\frac{1+\mu}{\mu}}$ and $\sqrt{\text{Var} (M_n) } \sim b(\ln n)^{\frac{1}{\mu}}\equiv\frac{a(1+\mu)\sqrt{6}}{\pi \mu} (\ln n)^{\frac{1}{\mu}}$ \cite{Majumdar:2020}, the probability $\mathbb{P} \left(\frac{M_n-a (\ln n)^{1/\mu+1}}{b (\ln n)^{1/\mu}} \leq x \right)$ converges to a Gumbel distribution $\Xi_\infty(x)$ with zero mean and unit variance. Consequently, $\mathbb{P}(N_{\mathcal{S}=x b (\ln n)^{1/\mu}+a (\ln n)^{1/\mu+1}} \geq n)$ also converges to $\Xi_\infty(x)$ for large $n$. By inverting $\mathcal{S}=x b (\ln n)^{1/\mu}+a (\ln n)^{1/\mu+1}$ in the limit of large $n$, we obtain:
	\begin{equation}
		\label{eq:NSTrans}
		\mathbb{P}(N_{\mathcal{S}}/\exp\left[ (\mathcal{S}/a)^{\mu/(1+\mu)}\right] \geq x) \sim \Xi_\infty\left(-\frac{\sqrt{6}}{\pi} \ln x\right)=\exp\left[ -x \E^{-\gamma_E} \right] \; .
	\end{equation}

	This implies that $N_\mathcal{S}$ asymptotically follows a single-parameter scaling: both its average and standard deviation grow exponentially with $\mathcal{S}^{\mu/(1+\mu)}$. Moreover, the asymptotic distribution of $N_\mathcal{S}$ is exponential, akin to the infinite-dimensional case (refer to Eq.~\eqref{eq:NS_tau_InfDim}). Finally, by taking the limit $\mu\to \infty$ in the expression for the average of $N_\mathcal{S}$, we recover the exponential growth of the average number of visited sites at starvation with the metabolic time, as obtained for the infinite-dimensional case in Eq.~\eqref{eq:NS_av_InfDim}.
\end{itemize}
The average, standard deviation and asymptotic distribution of $N_\mathcal{S}$ in the recurrent, marginal and transient cases are checked numerically in Fig.~\ref{fig:DistribN} for representative RWs of each class.

\begin{important}{Important}
	By considering the rescaled number of sites visited at starvation $\frac{N_\mathcal{S}}{\left\langle N_\mathcal{S} \right\rangle }$, where the asymptotics of $\left\langle N_\mathcal{S} \right\rangle$ is given by
	\begin{equation} \label{eq:momNS}
		\left\langle N_\mathcal{S} \right\rangle\sim \begin{cases}
			\mathcal{S}^\mu & [\mu<1] \\
		   \mathcal{S}^2 & [\mu=1] \\
		\exp\left[\mathcal{S}^{\mu/(1+\mu)}\right]	& [\mu>1] ,
		\end{cases}
	\end{equation}
	we obtain its asymptotic tail distribution $\mathbb{P}\left(\frac{N_\mathcal{S}}{\left\langle N_\mathcal{S} \right\rangle } \geq x \right)=\Phi_\mu(x)$, whose behavior at large and small arguments is summarized in Table~\ref{tab:NS}.
	\begin{table}[h!]
		\centering	
		\caption{Asymptotic properties of the minus log tail distribution $\Phi_\mu$ of the rescaled number of sites visited at starvation for different type of processes. All constants have been put to one for simplicity.}	
		\label{tab:NS}
		\begin{tabular}{ccc}
			\hline\noalign{\smallskip}
			$-\ln \Phi_\mu(x) $ & $x\ll 1$ & $x\gg1$   \\
			\noalign{\smallskip}\svhline\noalign{\smallskip}
			$\mu<1$ [recurrent]& $\E^{-1/x^{1/\mu}}/x^{1/\mu}$ &  $x$     \\
			$\mu=1$ [marginal]& $\E^{-1/x^{1/4}}$ &  $x$   \\
			$\mu>1$ [transient] & $x$ & $x$  \\
			\noalign{\smallskip}\hline\noalign{\smallskip}
		\end{tabular}
	\end{table}
\end{important}

\subsection{Lifetime}
\label{sec:lifetime}
We proceed by demonstrating that the distribution of $N_\mathcal{S}$ gives access to the distribution of the lifetime $T_\mathcal{S}$.\\

The lifetime $T_\mathcal{S}$ is determined by the accumulation of inter-visit times $\tilde{\tau}_k$ ($k<N_\mathcal{S}$), each corresponding to $\tau_k$ conditioned on being less than the metabolic time $\mathcal{S}$. Expressing the distribution of $T_\mathcal{S}$ yields:
\begin{eqnarray}
	&&\mathbb{P}(T_\mathcal{S}=t) \nonumber \\
	&&=\int_0^\infty \D n \mathbb{P} \left( \sum_{k=0}^{N_\mathcal{S}-1} \tau_k + \mathcal{S} = t| N_\mathcal{S} = n \right) \mathbb{P}(N_\mathcal{S} = n) \nonumber \\
	&& \approx \int_0^\infty \D n \delta \left( n\left\langle \tilde{\tau}_\infty \right\rangle + \mathcal{S} - t \right) \mathbb{P}(N_\mathcal{S} = n). \label{eq:TandNstarv}
\end{eqnarray}
In this derivation, we exploit the fact that for large $k$, the distribution of $\tilde{\tau}_k$ becomes independent of $k$, and the sum follows the law of large numbers, $\sum\limits_{k=0}^{n-1}\left\langle \tilde{\tau}_k \right\rangle \sim n \lim\limits_{k \to \infty} \left\langle\tilde{\tau}_k \right\rangle=n \left\langle\tilde{\tau}_\infty \right\rangle$. This leads to the tail distribution:
\begin{equation}
	\label{eq:TandNgeq}
	\mathbb{P}(T_\mathcal{S} \geq t) \approx \mathbb{P}\left(N_\mathcal{S} \geq \frac{t-\mathcal{S}}{\left\langle\tilde{\tau}_\infty \right\rangle} \right) \; .
\end{equation}

\begin{figure}[th!]
	\centering
	\includegraphics[width=\columnwidth]{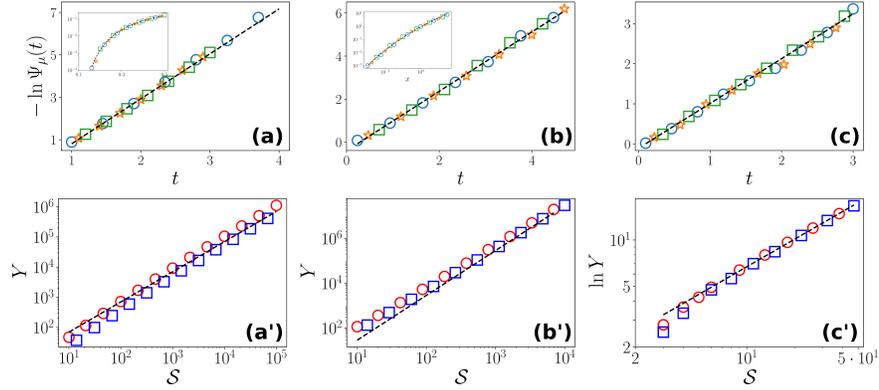}
	\caption{{\bf Lifetime in Starving Random Walks.} {\bf (a)-(c)} Distributions of the rescaled variable $t\equiv T_\mathcal{S}/\langle T_\mathcal{S}\rangle$ (insets illustrate behavior at small $t$-values), and {\bf (a$^\prime$)-(c$^\prime$)} corresponding averages (red circles) and standard deviations (blue squares) of $T_\mathcal{S}$. The black dashed lines denote the best fit from theory  (see Eqs.~\eqref{eq:RecLifetimeDist}, \eqref{eq:MargLifetimeDist}, \eqref{eq:TransLifetimeDist}, and \eqref{eq:momTS}). Various universality classes are depicted: {\bf (a)} $1d$ L\'evy flights with parameter $\alpha=1.5$ (recurrent), $\mathcal{S} = 14667$, $31622$, and $68129$; {\bf (b)} nearest-neighbor $2d$ random walks, $\mu=1$ (marginal), $\mathcal{S}= 2335$, $4832$, and $10^4$; {\bf (c)} nearest-neighbor $3d$ random walks, $\mu=3/2$ (transient), $\mathcal{S}=12$, $17$, and $22$. Increasing values of $\mathcal{S}$ are denoted by blue circles, orange stars, and green squares.}
	\label{fig:Lifetime}
\end{figure}

Then, we proceed with the characterization of $\tilde{\tau}_\infty$, which varies for each type of RW.
\begin{itemize}
	\item For recurrent RWs: At early times, the distribution $F_k(\tau)$ of inter-visit time $\tau_k$ behaves as $F_k(\tau)\propto\tau^{-1-\mu}$ (see Table.\ref{tab:Summary}). Consequently, the average value of $\tilde{\tau}_\infty$ scales as:
	\begin{equation}
		\left\langle \tilde{\tau}_\infty \right\rangle \propto \frac{\sum_{\tau=1}^{\mathcal{S}}\tau^{-\mu}}{\sum_{\tau=1}^{\mathcal{S}} \tau^{-1-\mu}} \sim c \mathcal{S}^{1-\mu}
	\end{equation}
	where $c$ is a model-dependent constant. Since $0<\mu<1$, the scaling $\mathcal{S}^{1-\mu}$ arises from the numerator. By rescaling $T_\mathcal{S}$ by $\mathcal{S}$ and using Eq.~\eqref{eq:NSRec}, we find:
	\begin{equation}
		\mathbb{P}(T_\mathcal{S}/\mathcal{S} \geq t)=\mathbb{P}\left(N_\mathcal{S} \geq c (t-1) \mathcal{S}^\mu \right)=\Xi_\mu((c(t-1))^{-1/\mu})
		\label{eq:RecLifetimeDist}
	\end{equation}
	where $\Xi_\mu$ is defined in Eq.~\eqref{eq:RecScale}. This reveals that $T_\mathcal{S}$ scales with $\mathcal{S}$ irrespective of the value of the exponent $\mu<1$, generalizing the $1d$ scenario (see Eq.~\eqref{eq:TS_av_formula_1d}). 
	\item For marginal RWs: Analogously to recurrent RWs, the average value of $\tilde{\tau}_\infty$ is given by:
	\begin{align}
		\left\langle \tilde{\tau}_\infty \right\rangle \propto \frac{\sum_{\tau=1}^{\mathcal{S}}\tau^{-1}}{\sum_{\tau=1}^{\mathcal{S}} \tau^{-2}} \sim c \ln \mathcal{S} \; .
	\end{align}
	Hence, by scaling $T_\mathcal{S}$ by $\mathcal{S}^2 \ln \mathcal{S}$, we obtain:
	\begin{equation}
		\label{eq:MargLifetimeDist}
		\mathbb{P}(T_\mathcal{S}/(\mathcal{S}^2\ln \mathcal{S}) \geq t)=\mathbb{P}\left(N_\mathcal{S} \geq t \mathcal{S}^2/c \right)=\Xi_1(\sqrt{c/t})
	\end{equation}
	resulting in a scaling form for the distribution of $T_\mathcal{S}$.
	\item For transient RWs: In this case, the inter-visit times' distribution has finite moments, leading to $\left\langle \tilde{\tau}_\infty \right\rangle \to \left\langle \tau_\infty \right\rangle= c$. By rescaling $T_\mathcal{S}$ by $\exp\left[ (\mathcal{S}/a)^{\mu/(1+\mu)}\right]$, we derive:
	\begin{eqnarray}
		\label{eq:TransLifetimeDist}
		&&\mathbb{P}\left(T_\mathcal{S}/\exp\left[ (\mathcal{S}/a)^{\mu/(1+\mu)}\right] \geq t\right)\sim \mathbb{P}\left(N_\mathcal{S} \geq t \exp\left[ (\mathcal{S}/a)^{\mu/(1+\mu)}\right] / c \right) \nonumber \\
		&&\sim \Xi_\infty\left(-\frac{\sqrt{6}}{\pi} \ln (t/c)\right)=\exp\left[-\frac{t}{c}\E^{-\gamma_E} \right]
	\end{eqnarray}
	where $a$ is the same constant as in Eqs.~\eqref{eq:def_m} and \eqref{eq:NSTrans}. The lifetime $T_{\mathcal{S}}$ follows an exponential distribution, akin to the number $N_{\mathcal{S}}$ of sites visited at starvation, as observed in the infinite-dimensional scenario (see Eq.~\eqref{eq:AvTSInfD}).
\end{itemize}
The average, standard deviation and asymptotic distribution of $T_\mathcal{S}$ in the recurrent, marginal and transient cases are checked numerically in Fig.~\ref{fig:Lifetime} for representative RWs of each class.

\begin{important}{Important}
	By considering the rescaled lifetime $\frac{T_\mathcal{S}}{\left\langle T_\mathcal{S} \right\rangle }$, where the asymptotic of $\left\langle T_\mathcal{S} \right\rangle$ is given by
	\begin{equation} \label{eq:momTS}
		\left\langle T_\mathcal{S} \right\rangle\sim \begin{cases}
			\mathcal{S} &[\mu<1] \\
			\mathcal{S}^2 & [\mu=1] \\
			\exp\left[\mathcal{S}^{\mu/(1+\mu)}\right]	& [\mu>1]  \; ,
		\end{cases}
	\end{equation}
	we obtain the asymptotic tail distribution $\mathbb{P}\left(\frac{T_\mathcal{S}}{\left\langle T_\mathcal{S} \right\rangle } \geq t \right)=\Psi_\mu(t)$, which has the properties at large and small arguments summarized in Table~\ref{tab:TS}.
	\begin{table}[h!]
		\centering	
		\caption{Asymptotic properties of the minus log tail distribution $\Psi_\mu$ of the rescaled lifetime for different type of processes. All constants have been put to one for simplicity.}	
		\label{tab:TS}
		\begin{tabular}{ccc}
			\hline\noalign{\smallskip}
			$-\ln \Psi_\mu(t) $ & $x\ll 1$ & $x\gg1$   \\
			\noalign{\smallskip}\svhline\noalign{\smallskip}
			$\mu<1$ [recurrent]& $\E^{-1/t^{1/\mu}}/t^{1/\mu}$ &  $t$     \\
			$\mu=1$ [marginal]& $\E^{-1/t^{1/4}}$ &  $t$   \\
			$\mu>1$ [transient] & $t$ & $t$  \\
			\noalign{\smallskip}\hline\noalign{\smallskip}
		\end{tabular}
	\end{table}
\end{important}

\subsection{Discussion}
In this section, we finally bridged the gap between the unidimensional and infinite dimensional starving RWs studied in Sec.~\ref{sec:LimitDimStarv}. 
\begin{svgraybox}
	Several important properties stem from the analysis of the general Markovian starving RW:
	\begin{enumerate}
		\item The number of sites visited at starvation grows algebraically with the metabolic time $\mathcal{S}$ for recurrent ($N_\mathcal{S}\propto \mathcal{S}^\mu$) and marginal ($\propto \mathcal{S}^2$) RWs, while it grows significantly faster for transient RWs ($\propto \exp\left[\mathcal{S}^{\frac{\mu}{1+\mu}}\right]$).
		\item The average lifetime grows linearly with $\mathcal{S}$ for recurrent RWs ($T_\mathcal{S}\propto \mathcal{S}$), while it grows quadratically for marginal RWs ($\propto \mathcal{S}^2$) and stretched exponentially for transient RWs ($\propto \exp\left[\mathcal{S}^{\frac{\mu}{1+\mu}}\right]$).
		\item The distribution of both the number of sites visited at starvation and the lifetime are scale-invariant, in the sense that $N_\mathcal{S} /
	\left\langle N_\mathcal{S} \right\rangle$ and $T_\mathcal{S} /\left\langle T_\mathcal{S} \right\rangle$ are independent of $\mathcal{S}$ for large $\mathcal{S}$. Additionally, both asymptotic distributions of the rescaled variables decay exponentially at large argument.
	\end{enumerate}
\end{svgraybox}

\section{Conclusion}

To conclude, we have shown how to derive key properties of the starving random walk (RW) problem, such as the number of sites visited at starvation ($N_\mathcal{S}$) and the lifetime of the RW ($T_\mathcal{S}$). We have demonstrated that understanding the universal properties of the time between visits ($\tau_k$) leads to universal characteristics for these quantities for general symmetric Markovian RWs. We have highlighted important differences based on a single parameter $\mu$, which characterizes recurrence or transience of RWs. For example, exploration in one or two dimensions (i.e., by recurrent walks) differs significantly from that in three dimensions (i.e., by transient walks), where the lifetime of a tracer with a specific metabolism time changes from algebraic (for recurrent walks) to stretched exponential (for transient walks).


This chapter's methods and findings serve two purposes:\\ First, they provide a basic framework for studying systems with limited resources, useful in fields like animal ecology \cite{orlando2020power} or the study of biological species in crowded environments \cite{Passino:2012}. In these situations, significant memory effects arise because the RW is influenced by its past path, which is affected by resource depletion.


Second, our results furnish a null model against which the exploration dynamics of biological tracers can be assessed. Departures from this model indicate the potential influence of intricate and realistic strategies not encompassed in this study, including the tracer's memory of its trajectory (using biological cues, such as ants or MDCK cells do \cite{Gordon:1995,Alessandro:2021}), renewal of the resources with time \cite{Chupeau:2016} or the option for the tracer to abstain from consuming food when unnecessary \cite{Benichou:2018}. Other extensions include the food detection at long distances \cite{sanhedrai2020lifetime,Sanhedrai2021}, penalized long moves \cite{Krishnan2018}, the intermittent motion \cite{campos2021optimal}. The starving RW problem provides a relatively straightforward model yielding numerous insights and results, with the methods presented in this chapter laying the foundation for a comprehensive understanding of resource-limited motion in a broad range of systems.

\bibliographystyle{spphys}
\bibliography{biblio}

\end{document}